# Initial PIP-II Beam Current Monitor Fault Case Analyses & Beam Position Monitor Linearity Studies in CST Studio Suite

A. R. Rouzky†, Duke University, Durham, NC, 27708, USA
N. Eddy, M. A. Ibrahim, Fermilab, Batavia, IL, 60510, USA

*Abstract*

The use of non-invasive sensors & systems to measure particle beam characteristics is a crucial part of modern accelerator control systems due to their ability to return real time passive measurements without impacting the beam quality. Simulations, which can predict these sensors' behaviour and performance under anticipated accelerator conditions, are valuable tools to ensure confidence in the sensors' functionality prior to a physical bench test. This paper details the design, testing, and results of two sensor models developed using CST Studio Suite software. One model is an elliptical, large-aperture beam position monitor (BPM) for which vertical & horizontal position signal linearity was analyzed. The second model is an AC current transformer (ACCT) beam current monitor (BCM), which was used to search for potential fault cases within the BCM and beam pipe flange gaps. Fermilab Proton Improvement Plan II (PIP-II) accelerator beam conditions were applied, and special focus is given to the discovery of linearity variations within the BPM as well as the use of frequency domain techniques in the BCM fault case analyses.

## INTRODUCTION

Due to the high precision requirements of modern high energy physics experiments, it is of critical importance that current accelerators possess a high degree of control stability and modularity. This precipitates the necessity of high-accuracy beam diagnostics, which, in turn, influences the physical design of current monitors (BCMs) and beam position monitors (BPMs), to satisfy stringent physics requirements specific to an arbitrary experiment.

Fermilab's PIP-II will enable the world's most intense neutrino beam and give scientists unique capabilities to study rare particle physics processes. Upon completion, PIP-II will accelerate protons at up to 800-million-electronvolts, or 800 MeV, over its 215-meter length, with an instantaneous beam power of more than 1 megawatt.

PIP-II's beam line lattice incorporates models for various beam instrumentation sensors. While mechanical models have been integrated to confirm general physical compatibility with the beamline locations and geometry, further simulations were needed to examine sensor performance at those locations. Specifically, this work was developed to investigate PIP-II large aperture BPMs as well as the PIP-II ACCT BCMs, with customized flanging.

PIP-II large aperture BPM simulations focused on the difference-over-sum position signal linearity for various beam positions within an elliptical-plate BPM. The findings, in turn, would allow mechanical designers to fine-tune the geometry of the physical design to provide the appropriate signal dynamic range for the data acquisition electronics.

Driven by recommendations from a PIP-II Final Design Review (FDR) committee, the PIP-II ACCT simulations were motivated by concerns regarding arcing experienced on errant beam pulses within the Spallation Neutron Source at Oak Ridge National Laboratory [1]. BCM arcing was examined throughout portions of the PIP-II beamline by varying the BCM geometry and beam parameters to match the conditions at a desired location along the PIP-II beamline and measuring the induced flange gap voltages.

## SIMULATION METHODS

The simulations were performed in CST Studio Suite, a high-performance 3D electromagnetic analysis software package for designing, analyzing and optimizing electromagnetic (EM) components and systems. EM field solvers for applications across the EM spectrum are contained within a single user interface in CST Studio Suite software. Although both models used CST Studio's wakefield solver, the methods used within this work differed depending on the sensor type, as the simulation goal (beyond simply developing functioning models of the sensor) for each analysis was different.

### BCM Fault Case Analyses

Throughout the PIP-II beamline, in-flange Bergoz ACCT units shall be installed as the primary sensor for performing current measurements of transferred beam pulses. In-house zero-length flanges will be used to adapt the ACCT's flanges to the beamline flanges. Potential vacuum arcing at these flanges due to unwanted electromagnetic resonances was examined using the BCM simulations detailed in this work. Such arcing would not only risk inducing undesired perturbations in the beam, but also pose a significant general safety risk to any personnel within the vicinity of the arcing flanges.

Using the provided mechanical schematics of both the ACCT and the flanges, a BCM model was created in CST. The model was parameterized to facilitate varying the BCM geometry and beam parameters at various locations throughout PIP-II. Figure 1 shows the cross-section of the BCM model, adjusted for the physical dimensions of the sensor to be installed in the PIP-II warm front end (WFE).

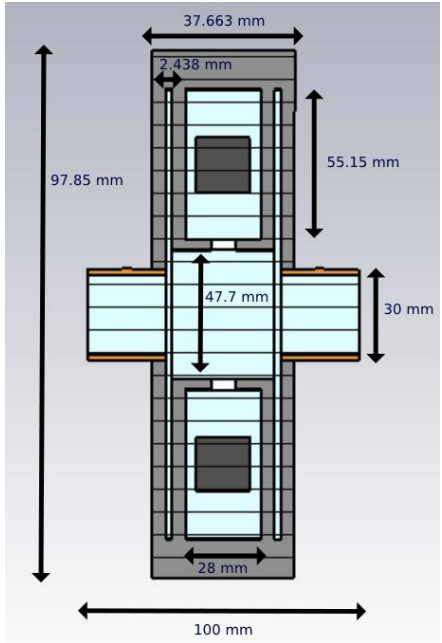

Figure 1 : Cross-section of the parametric BCM model used in the fault case analyses, including measurements. Geometry was kept consistent for each run, except for beam pipe diameter which varied from 30 to 50 mm.

For simplicity, the BCM and its connecting flanges to the beam pipe (gray) were simulated as a single block of a perfect electric conductor (PEC), with the flange gaps taking the form of vacuum cylinders cutting through the object. The beam pipe (orange) was also simulated as a PEC. The iron core (black) was simulated as a material with a frequency-dependent magnetic permeability based off FT-3KM nanocrystal (Figure 2).

The BCM ceramic gap (visible as a white block underneath the core) was simulated as a perfect electric insulator with the same permittivity and permeability as a perfect vacuum ($\mu_r = \varepsilon_r = 1$). The remaining blue sections were all additionally simulated as a perfect vacuum. While it is technically true that the region surrounding the nanocrystalline core would be air, and not a perfect vacuum, the effects of this on the presence of flange arcing would likely be minimal, and as such this discrepancy is tolerated.

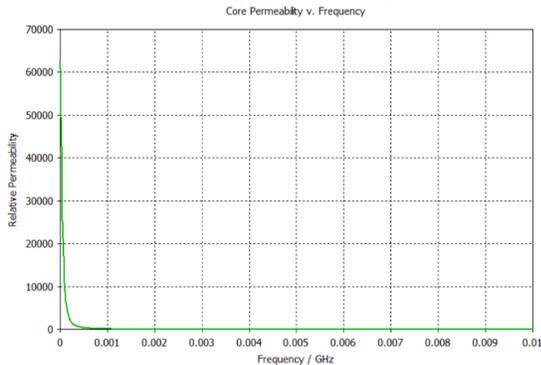

Figure 2 : Graph of the frequency-dependent relative magnetic permeability $\mu_r$ of the nanocrystal. Values taken from Hitachi Metals [2].

The majority of the BCM simulation runs used a centered particle beam, so to reduce simulation time and take advantage of the symmetry of the mesh, the XZ and XY planes were set as magnetic symmetry planes. The beam pipe was simulated as an open port on both ends ($X_{min}$ & $X_{max}$); all other boundary conditions were set as $E_t = 0$.

To properly simulate the pulsed beam bunches that would pass through the structure and potentially give rise to arcing in the flange gaps, a custom VBA excitation signal (Figure 3) was needed to generate a train of gaussians pulses. Each gaussian, with standard deviation of σ, was separated by a period determined by the PIP-II beam frequency (169.5 MHz).

However, due to CST Studio's restriction on simulating custom excitation signals for non-ultra-relativistic beams ($\beta < .999$), direct simulation of this signal was not possible. To circumvent this, frequency domain techniques were applied using CST's post-processing features.

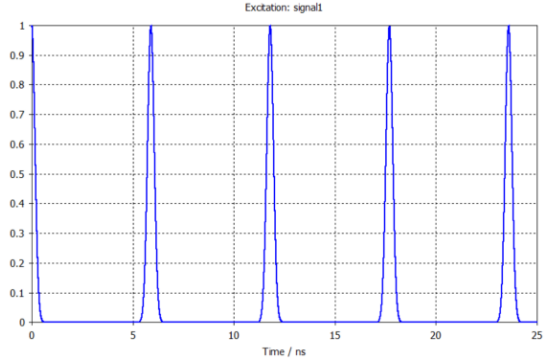

Figure 3 : A 25ns sample of the VBA excitation signal for $\beta = .067$, containing four full gaussian bunch pulses at 169.5 MHz.

First, a single gaussian bunch (σ = 6.6 mm) passing through the structure was simulated using CST's default gaussian excitation signal, with simulation $F_{min} = 0$ Hz & $F_{max}$ as:

$$F_{max} = \left(\frac{3\sigma}{\beta * c}\right)^{-1}$$

where c is the speed of light.

The effects of the bunch on the flange gap voltages were measured for the duration of one entire beam pulse (which would range from 10μs to 500μs) using voltage monitors. Afterwards, a Fourier transform (FT) was used to find the frequency spectrum of the flange voltages.

Next, the CST Fourier series (FS) result template step was used to transform the custom excitation signal into an infinitely repeating gaussian train at 169.5 MHz. To perform these operations on the excitation signal with post-processing, it was exported and re-imported as a 1D Result. The product of the gaussian train FS and the flange voltage FT was then taken (thereby convolving the two in the time domain) to find the flange voltage output spectrum. An inverse Fourier transform was then performed on this output spectrum to transform it into the time-domain, providing an approximation of the time-domain flange voltages created by an arbitrarily long train of gaussian particle bunches. This process of taking advantage of the frequency domain to analyze periodic signals is recommended and further detailed in CST Studio official documentation [3].

*BPM Linearity Studies*

Within the PIP-II beamline, large aperture BPMs will be installed near the Lambertson septum magnet, at the junction between the Booster Absorber line and the final transfer line arc to the Booster. These BPMs sensors, which use an in-house shoebox design, must accommodate the 160mm x 40mm elliptical beampipe of that section, and accurately provide position measurements to determine the beam trajectory. CST simulations were used to perform linearity studies on these sensors by simulating the response of the sensor to an incident particle beam. Table 1 lists the parameters of the particle beam source, which were kept constant for each simulation run.

Table 1 : Beam Parameters Used for BPM Simulations

| Parameter Name | Parameter Value |
|---|---|
| B | .8 |
| σ (mm) | 150 |
| Charge (C) | 1.2422 x $10^{-11}$ |

Initial simulations were performed using an imported NX Studio Model of a shoebox BPM, which formed an approximation of an elliptical BPM with a plate gap using two split square plates rather than two elliptical plates. However, the linearity studies detailed in this paper were performed with a CST model based on a split-plate elliptical BPM, previously used at the Facility for Rare Isotope Beams (FRIB) at Michigan State University. Figure 4 shows a cross-section and isometric view of this model. The elliptical plates & guard ring (gray) are simulated as PEC, with the rest of the region as vacuum (blue).

The data used to model BPM signal linearity was obtained with voltage monitors to measure the time-dependent voltage signals of the BPM's elliptical plates, which would vary depending on the vertical and horizontal positions of the simulated particle beam. The terms "horizontal" and "vertical" refer to the BPM's z and y-axis, respectively.

The beam itself was simulated as a single gaussian bunch using CST Studio's default gaussian excitation signal, with simulation $F_{min}$ & $F_{max}$ changing depending on the desired mesh fineness of the simulation (as CST Studio accounts for the maximum frequency of a simulation when calculating mesh cell count). All full-range position sweep simulations were performed for a wakelength of 2000 mm. Boundary conditions were set as $E_t = 0$ for all direction minima and maxima, and no symmetry planes were used.

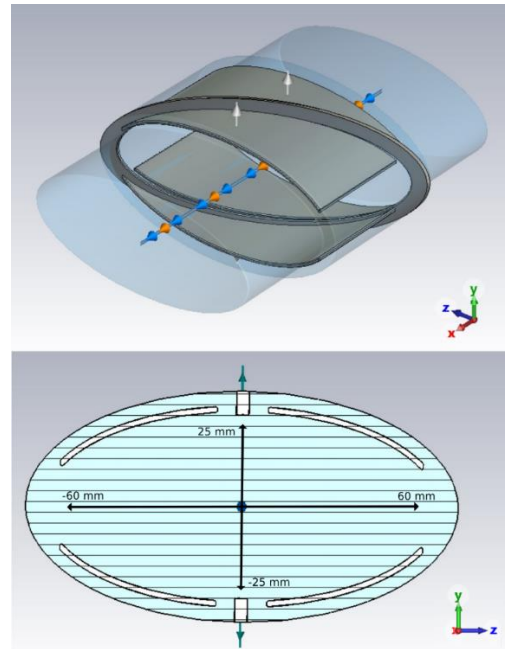

Figure 4 : Cross-section and isometric view of the primary BPM model used for linearity simulations, including annotations for vertical and horizontal beam sweep ranges. Voltage monitors & centered particle beam are visible.

Using CST Studio's parametric sweep feature, the beam's position was varied such as to take it from one edge of the BPM to the other, both horizontally and vertically. Horizontal and vertical sweeps were performed in isolation, i.e. during a parametric sweep of the beam's vertical position (y=-25 mm to y=25 mm), the beam maintained a centered horizontal position (z = 0 mm) and vice versa.

After a parametric sweep, CST post-processing was used to integrate the voltage monitor time signals of the four plates over entire duration of the signal, a process which was performed for each beam position in the sweep. This provided a function of plate voltage integral vs. beam position, which depicted how the signal of a single plate would vary through a beam position sweep. Figure 5 shows an example of the resulting measurement.

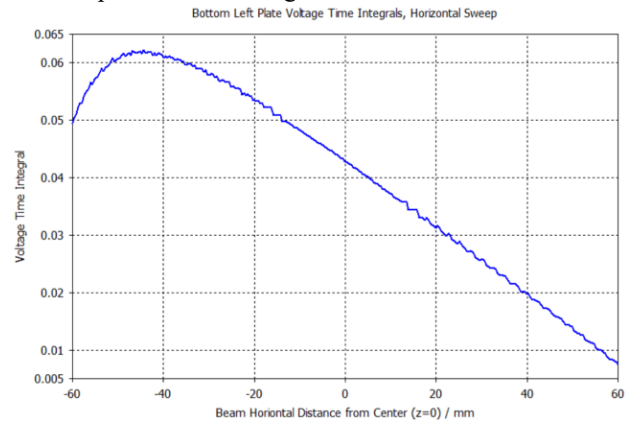

Figure 5 : Voltage time integrals of the bottom left (-z, -y) elliptical plate for a full horizontal beam position sweep.

The difference-over-sum normalized position signal was then acquired using CST Studio's "Mix 1D Siginals" function and the following equation:

$$\frac{(A + B) - (C + D)}{A + B + C + D}$$

Where the values of $(A + B)$ and $(C + D)$ were dependent on the direction of the beam position sweep.
- For a vertical sweep, $(A + B)$ would be the sum of two top plates, while $(C + D)$ be the sum of the two bottom plates.
- For a horizontal sweep, $(A + B)$ would be the sum of the two left plates, and $(C + D)$ would be the sum of the two right plates.

This provided two graphs of the normalized position signal, one per sweep direction. In addition, to better visualize the signal linearity, CST post-processing was once again used to find and plot the point-to-point derivative of the two position signals.

## SIMULATION RESULTS

### BCM Fault Case Analyses

Time signal voltages for a beam pulse obtained as detailed in the methods section universally returned lower flange voltages than expected. However, since this amplitude was found to be heavily dependent on the number of samples taken by the IFT, and the fact that CST post-processing returns a perfectly reconstructed signal when taking the IFT (with default settings) of an unmodified FT of an arbitrary signal, it is believed that this issue may be caused by incorrect frequency sampling that leads to errors in the final voltage output spectrum product. Figure 6 and Figure 7 highlight the differences in magnitude between the time-domain voltage signals of the left flange-gap for a single, $\beta = .841$ gaussian beam pulse, and the reconstructed signal for a train of said bunches at 169.5 MHz.

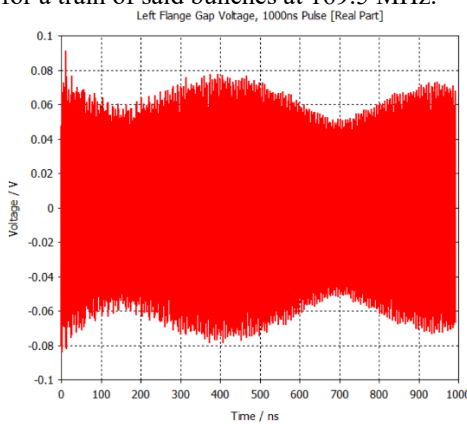

Figure 6 : Time-domain reconstruction of the left-flange gap voltage for a 1000 ns pulse. Note the lower magnitudes with respect to Figure 7.

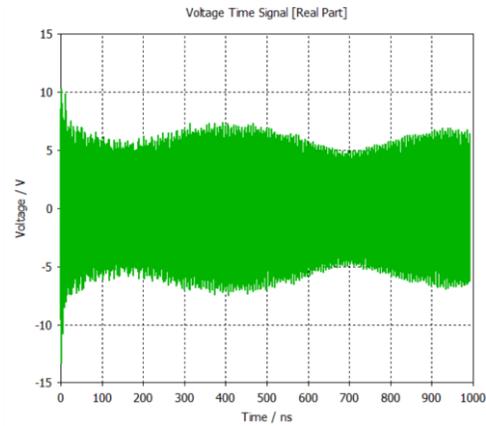

Figure 7 : Time-domain reconstruction of the original gap voltage signal simulated for a single pulse. Note the higher magnitudes with respect to Figure 6.

### BPM Linearity Studies

As shown in Figure 8, initial simulations returned signals that, at certain positions, behaved rather erratically. This was especially true for horizontal sweeps, which showed extreme signal variation around +/- 14-15 mm from the BPM center.

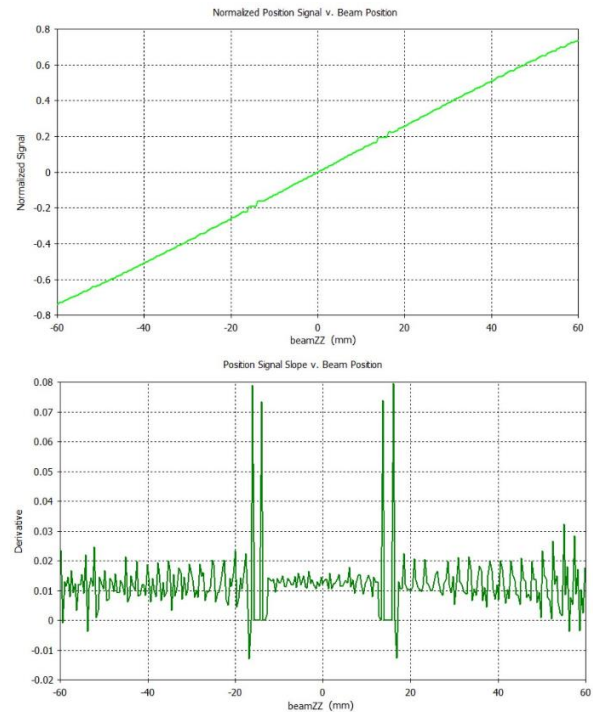

Figure 8 : Normalized BPM position signal & its slope for a full horizontal sweep. Note the prominent noise at +/- ~14-15 mm, visible as spikes in the slope graph.

It was initially believed this may be due to insufficient mesh fineness, as the mesh seemed to possess wider gaps at the problematic areas as well as a relatively low cell count $c_{count}$ of 120,960. However, further simulations performed around the problematic coordinates (z=14-17 mm) with a mesh of significantly higher fineness ($c_{count}$ =

8,047,152) returned linearity variations of a similar magnitude. It should be noted that, due to time constraints, this limited-scope sweep of the finer mesh was only simulated for a wakelength of 1200 mm (8σ). Table 2 provides a comparison between the initial and refined meshes. Additionally, Figure 9 compares the resulting position signal slope for each of the mesh settings investigated.

Table 2 : BPM Model Meh Settings

| Simulation Property | Initial "Coarse" Mesh | Refined Mesh |
| --- | --- | --- |
| $f_{min}$ (GHz) | 0 | 0 |
| $f_{max}$ (GHz) | 1 | 6 |
| Minimum Cell Size (x, y, z) | 1 | .7 |
| Cells per max model box edge | 10 | 40 |
| Cells per wavelength | 10 | 50 |

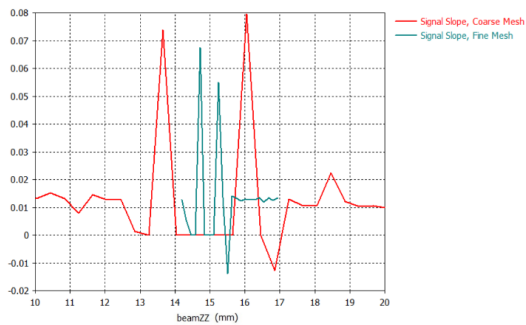

Figure 9 : Overlaid plots of the position signal slope for the coarse and fine meshes.

## CONCLUSIONS

Frequency domain analyses have the potential to be a highly effective tool for the simulation of periodic gaussian bunches within a larger beam pulse. Further work is required to determine how to extract a physically valid voltage time signal from the flange voltage output spectrum. This may involve performing certain post-processing steps (such as the frequency-domain multiplication and final IFT) using alternative software, such as MATLAB, to design more customizable post-processing scripts to avoid pitfalls that may be inherent to the methods CST uses to perform these operations.

The large-aperture BPM model has proven to be a highly useful tool for validating performance of the sensor before beamline implementation. However, further analyses on the simulated BPM position signals are required to determine the causes of the aforementioned noise. Current results seem to indicate that insufficient mesh fineness is not the cause of the issue, although other aspects of the simulation should be reviewed to determine if these results are due to physical issues with the BPM geometry, simple quirks of CST Studio's numerical analysis, or both.

## ACKNOWLEDGEMENTS

The author would like to thank the entire Beam Instrumentation department for their willingness to assist this work in any way they can. In particular, they thank Randy Thurman-Keup for his assistance throughout various CST Studio roadblocks for both the BCM and BPM. The author would also like to thank Dakota Krokosz for creating the first NX Studio BPM model and helping the author during their first experiences with CST Studio, and Jenny Crisp for providing the BPM model that was ultimately used in the BPM linearity studies.